\documentclass[article,a4paper,oneside,hyphens]{memoir}

\usepackage[T1]{fontenc}
\usepackage{lmodern}
\usepackage{url}
\usepackage{multicol}
\usepackage{latexsym}
\usepackage{listings}
\usepackage{hyperref}

\usepackage{vmargin}
\setpapersize{A4}

\newcommand{\published}[1]{%
    \gdef\puB{#1}}
 \newcommand{\puB}{}

\newcommand{\code}[1]{\texttt{\footnotesize #1}}

\setsecnumdepth{subsubsection}

\published{Originally published in
	\textit{BSD Magazine}, june 2014 issue --
	\url{http://bsdmag.org}
	\thanks{Reprinted with permission}}
			  
\title{TLS hardening}
\author{Emmanuel Dreyfus}

\begin{document}

\maketitle

\begin{abstract}This document presents TLS and how to make it secure enough 
as of 2014 Spring. Of course all the information given here will rot with
time. Protocols known as secure will be cracked and will be replaced
with better versions. Fortunately we will see that there are ways to 
assess the current security of your setup, but this explains why you 
may have to read further from this document to get the up to date
knowledge on TLS security.

We will first introduce the TLS protocol and its underlying components:
X.509 certificates, ciphers, and protocol versions. Next we will have
a look at TLS hardening for web servers, and how to plug various
vulnerabilities: CRIME, BREACH, BEAST, session renegotiation, Heartbleed, 
and others. We will finally see how the know-how acquired on hardening
web servers can be used for other protocols and tools such as Dovecot, 
Sendmail, SquirrelMail, RoundCube, and OpenVPN. 

We assume you already maintain services that use TLS, and have basic 
TCP/IP network knowledge. Some information will also be useful for
the application developer.
\end{abstract}

\setlength{\parskip}{0.5\baselineskip}

\section{An introduction to TLS}
TLS stands for Transport Layer Security. It is an encryption and authentication
layer that fits between transport and application level in the TCP/IP 
network stack. It got specified by IETF in 1999 as an enhancement over 
Netscape's Secure Socket Layer (SSL), which is why we often see the SSL 
term used instead of TLS.

TLS is easy to add on top of any TCP service, and this is why it grown so
popular, and it became available for many protocols. For instance, 
HTTP can be used over TLS, using well-known \lstinline{https://} URL. It works
the same way for SMTP(S), IMAP(S), POP(S), LDAP(S), and so on.

\subsection{X.509 certificates}
The main goal of TLS is enforcing confidentiality and integrity.
This cannot happen if the remote peer is not properly authenticated:
who cares about having a secure channel if we do not know who we
are speaking to? Attacks where an intruder slips between the two
legitimate parties are known as Man In The Middle (MITM or MiM) attacks.
In such a setup, a secure channel exists between one legitimate
party and the intruder, and there is another secure channel between the
intruder and the second legitimate party.  The legitimate parties
talk to each others, and the intruder sees all the traffic.

TLS attempts to authenticate the remote party using a Public Key 
Infrastructure (PKI). The idea is to use asymmetric cryptography, 
where each party has a private key capable of performing cryptographic
signatures, and a public key, which can be used to verify a signature 
done by the associated private key. If a remote party is able to 
produce a public key validated signature for a nonce it was given,
that proves it has the private key.

We are therefore able to authenticate a machine for which we already 
know a public key. That leaves a problem to solve: how can we learn
the public key for a machine we never connected to?

Enter X.509 certificates, which are also known as SSL certificate. 
A X.509 certificate is a public key, with data such as the machine name
attached, and signed by a private key which is known as a Certificate
Authority (CA). When connecting to a server using TLS, the server sends
its X.509 certificate. Provided we have the public key of the CA that
signed it, we can check the signature and have a hint the public key
is the right one for the host we want to connect to, and nobody 
attempted to perform a MITM attack. Of course if the software has bugs, 
this check can be incomplete, and this is what happened to Apple \cite{1}
and GnuTLS \cite{2} recently.

But this mechanism relies on the following assumption: we have the
public key of the CA. How do we have it? For Unix command-line
tools such as \lstinline{wget} or \lstinline{curl}, you need to
install it first. This means it cannot work out of the box, but it
guarantees a level of security, as you can choose what CA you trust.
Note that some distributions provide \lstinline{curl} and
\lstinline{wget} configured with a system-wide CA repository, hence
your millage may vary. Try downloading a well known \lstinline{https://}
to discover the behavior of your system.

At the other end of the spectrum, we have web browsers and mail clients, 
which are bundled with the public keys for hundreds of commercial CA. 
Since any CA can sign a certificate for any machine name, this means
a web browser user trusts hundreds of CA, some of them running in 
oppressive jurisdictions where a government can compel the CA
to sign a certificate for someone else's machine. The CA can also
be compromised by hackers \cite{3}, with the same result of having someone
able to impersonate a machine during TLS authentication.

There are software to help defend against such attacks. For instance, 
Firefox has a Cert Patrol module that alerts you when the certificate
of a web site is not the same as usual. That will let the user spot 
a MITM attack using a rogue certificate, except of course when connecting
for the first time. 

There is an even worse attack possible on X.509 certificates, when
the cryptography used to build the certificate is too weak and can 
be cracked. For instance, a 512 bits RSA private key, which was once 
safe, is now vulnerable to factorization attacks \cite{4}, where the private 
key can be easily derived from the public key by computation. Once the
private key is discovered, an attacker can just passively record TLS traffic
and decipher it. We will see later that this can be avoided using 
Perfect Forward Secrecy (PFS), but even in that situation, an 
attacker can still decipher the traffic by running a MITM attack.

This leads us to the key length question: How long should it be?
The longer it is, the longer it will take before it can be compromised. 
But a key too long means slower cryptographic operations, and perhaps
software incompatibility. 1024 bits RSA is the next target, hence you 
should consider using 2048 bits RSA. 4096 bits RSA is even safer and 
will work most of the time, but there can be compatibility issues.

Huge key length is not always a guarantee for strong cryptography. Most
cryptographic operations need a random source, and if an attacker can
predict it, she may be able to decipher data or tamper with it. The private
key generation step critically needs a good random generator. If your 
operating system warns you about low entropy for random source, make sure
you fix it before generating private keys. Modern systems feed their 
random source from entropy gathered from various sources as they run. After
a reboot, the entropy pool is low, but the system may be configured to 
save and restore that information across reboots. In virtualized setups, 
the virtual machine may be configured to grab entropy on start up 
from the hypervisor.

Sometimes you will learn about a random generator weakness in your 
operating System. This can be the result of an implementation bug, 
which happened for instance to Debian \cite{5}, or a design mistake, for 
instance in the Dual Elliptic Curve random number generator. In the later case,
the mistake was made on purpose by the NSA, as a document leaked by 
former NSA contractor Edward Snowden taught us \cite{6}. In both situation, 
you first need to fix the random number generation process, then 
regenerate keys, and finally think about what traffic may have been 
compromised.

While we are on the key factoring problem, it is worth mentioning
the Shor algorithm, which allows quick factorization of decent 
sized keys. Fortunately this algorithm requires a quantum computer to
operate, and as of today we are not aware of any implementation of
such a system. But if it appears some day, asymmetric cryptography
as it is used in TLS will become almost as useless as a Caesar's code \cite{7}
is today.

We talked about RSA keys. RSA is indeed the most used public key
algorithm. It was invented by Rivest, Shamir and Adleman, hence the
name. There are alternatives: DSA, and ECDSA. The latter is not yet
widely supported, but it is interesting because it allows much
faster operations, which is very valuable on embedded devices.

\subsubsection{Key points}
\begin{itemize}
\item Keep software up to date to avoid running known bugs in certificate 
  validation and weak cryptography.
\item Make sure your certificate are signed by a CA known by the client.
\item Use certificate tracking software like Cert Patrol on the client.
\item Make sure your system random generator is not predictable.
\item Use RSA keys for compatibility, if possible, also use ECDSA keys.
\item Use key length as long as possible considering performance and 
  compatibility issues.
\end{itemize}

\subsubsection{How-to}

\paragraph{Create a 4096 bits RSA private key with appropriate file system 
   permissions}~

\begin{lstlisting}
  ( umask 077; openssl genrsa -out private.key 4096 )
\end{lstlisting}

\paragraph{Create a Certificate Signing Request (CSR, to be sent to a CA for signature)}~
\begin{lstlisting}
  openssl req -new -key private.key -out certificate.csr
\end{lstlisting}

\paragraph{Inspect a private key (beware the terminal will display private data)}~

\begin{lstlisting}
  openssl rsa -text -in private.key
\end{lstlisting}

\paragraph{Inspect a Certificate Signing Request}~
\begin{lstlisting}
  openssl req -text -in certificate.csr
\end{lstlisting}

\paragraph{Inspect a certificate (signed from the CSR by a CA)}~
\begin{lstlisting}
  openssl x509 -text -in certificate.crt
\end{lstlisting}

\subsection{Ciphers}

TLS uses various algorithms known as ciphers to check data integrity
and encrypt it.  The specifications make many ciphers available,
and implementations do not have to implement them all. The high
level of cipher choice means we will have to make sure the right
ones are used. This is important, because among ciphers, some are
secure for today's standards, and some are trivial to crack. There
are even null-ciphers, which perform no encryption at all. This is
useful for debugging, but of course it should not be used for
anything meant to be confidential.

The complete list of available ciphers for OpenSSL-based
software can be obtained by running the following command: 
\begin{lstlisting}
  openssl ciphers -v
ECDHE-RSA-DES-CBC3-SHA  SSLv3 Kx=ECDH     Au=RSA  Enc=3DES(168) Mac=SHA1
ECDHE-ECDSA-DES-CBC3-SHA SSLv3 Kx=ECDH     Au=ECDSA Enc=3DES(168) Mac=SHA1
SRP-DSS-3DES-EDE-CBC-SHA SSLv3 Kx=SRP      Au=DSS  Enc=3DES(168) Mac=SHA1
SRP-RSA-3DES-EDE-CBC-SHA SSLv3 Kx=SRP      Au=RSA  Enc=3DES(168) Mac=SHA1
EDH-RSA-DES-CBC3-SHA    SSLv3 Kx=DH       Au=RSA  Enc=3DES(168) Mac=SHA1
EDH-DSS-DES-CBC3-SHA    SSLv3 Kx=DH       Au=DSS  Enc=3DES(168) Mac=SHA1
ECDH-RSA-DES-CBC3-SHA   SSLv3 Kx=ECDH/RSA Au=ECDH Enc=3DES(168) Mac=SHA1
ECDH-ECDSA-DES-CBC3-SHA SSLv3 Kx=ECDH/ECDSA Au=ECDH Enc=3DES(168) Mac=SHA1
DES-CBC3-SHA            SSLv3 Kx=RSA      Au=RSA  Enc=3DES(168) Mac=SHA1
PSK-3DES-EDE-CBC-SHA    SSLv3 Kx=PSK      Au=PSK  Enc=3DES(168) Mac=SHA1
(...)
\end{lstlisting}

For the curious, each cipher has an IANA-registered number \cite{8} used 
in the TLS protocol. On recent versions of OpenSSL, that number can
be displayed using openssl ciphers -V

Cipher names, in the first column, contains hyphen-separated components. 
We find here the encryption algorithm, with optional key length. 
Here are a few examples:
\begin{description}
\item[DES]	The ancient Data Encryption Standard, with 56 bits keys
\item[3DES]	Triple DES, which is equivalent to a 168 bit keys
\item[RC2]	Ancient and insecure Rivest Cipher v2, with 40 bit keys
\item[AES128]	Modern Advanced Encryption Standard, with 128 bit keys
\item[AES256]	Modern Advanced Encryption Standard, with 256 bit keys
\end{description}

The key here is different from the private key used in X.509 certificate.
The latter uses asymmetric cryptography, which uses a lot of CPU power,
while the former uses symmetric cryptography, which costs much less 
processing. As the names suggests, symmetric cryptography uses a secret
shared by both parties as the key, which may be renegotiated at regular 
interval in time. Because cipher keys are of different nature, their
lengths are much shorter than X.509 certificates key lengths, but that
does not means they are insecure.

Then we find the hash algorithm. Here are a few examples:
\begin{description}
\item[MD5]	Ancient and now insecure Message Digest v5
\item[SHA]	Soon to be insecure Secure Hash Algorithm v1
\item[SHA384]	Modern SHA v2 with 384 bits long hash
\end{description}

The cipher name can also specify the X.509 certificate private key 
flavor, if the cipher belongs to a specification recent enough to 
support something else than RSA:
\begin{description}
\item[RSA]	Well known Rivest, Shamir and Adleman
\item[DSS]	Digital Signing Signature, used with DSA keys
\item[ECDSA]	Elliptic Curve DSA
\end{description}

There are different family of ciphers, and within a family, a given 
cipher may have different modes of operation. This may or may not be
reflected in the cipher name:
\begin{description}
\item[Stream ciphers] whose last usable TLS cipher is RC4 (Rivest Cipher v4)
\item[Block ciphers] which can operate in various modes: ECB, CBC, CTR, GCM...
The relevant modes for TLS block ciphers are:
\begin{description}
\item[CBC] Older Cipher block chain, which is quite common with TLS
\item[GCM] More secure Galois/Counter Mode, introduced with TLSv1.2. 
\end{description}
\end{description}

Within the block ciphers, CBC mode provides encryption without integrity.
This means a CBC mode cipher always needs a helper Hash Message 
Authentication Code (HMAC) function to enable integrity. GCM mode do not
have this requirement since it provides both encryption and integrity.

And finally, the cipher may negotiate its private keys using a 
Diffie-Hellman (DH) exchange:
\begin{description}
\item[EDH]	Diffie-Hellman (DH) Exchange
\item[DHE]	Ephemeral EDH, which enables PFS (see below)
\item[ECDH]	Newer and faster Elliptic Curve Diffie-Hellman
\item[ECDHE]	Ephemeral ECDH, which enables PFS (see below)
\end{description}

DH exchange support is an optional but very valuable feature. It
ensures that the symmetric cipher keys cannot be easily computed
from stored TLS exchanges if the X.509 certificate private key is
compromised in the future. That feature is called Perfect Forward
Secrecy (PFS), or just Forward Secrecy (FS) since nothing is perfect.
It is of no help for the ongoing TLS exchanges done after the X.509
certificate private key is compromised, but it protects stored
communications, which is interesting since we know the NSA stores
everything it can.

Some clients only implement ECDHE, hence it is desirable to support it
although it may require quite recent versions of softwares, as we will 
see later for Sendmail. For Apache, latest 2.2.x will support it. 
And while we deal with Elliptic Curve cryptography, it is important to 
note that the algorithm may use different constants on which the 
administrator has no control, and some values may be less safe than 
others \cite{9}, although we do not know practical attacks on that front yet.

Client and server negotiate a common cipher, which should obviously
be known by both implementations. On both side, the software may allow
an ordered cipher suite to be configured. For instance, Apache does 
this through the \lstinline{SSLCipherSuite} 
directive. The client setting usually
prevail, unless the server requests otherwise. This is achieved on 
Apache using the \lstinline{SSLHonorCipherOrder} directive.

In OpenSSL-based software the cipher suite syntax is a colon-separated
list of cipher names, with some additional syntax explained in
\lstinline{openssl_ciphers(1)} man page. Here are a few examples:
\begin{description}
\item[ECDH] selects all ECDHE-enabled ciphers
\item[HIGH] selects all high security ciphers (128 bit key length and beyond)
\item[TLSv1] selects all TLSv1 ciphers
\item[MD5] selects all ciphers using MD5
\end{description}

The \lstinline{openssl ciphers} command can be used to see all the 
ciphers enabled by a particular cipher suite. Here is an example
\begin{lstlisting}
  openssl ciphers ECDH:!RC4:!MD5:!NULL
\end{lstlisting}

A good cipher suite specification favors the high security ciphers, forbids 
the insecure, and allows enough ciphers so that any client can connect. Here
is a cipher suite specification that favors PFS and ciphers with 128 
bit keys and beyond, while remaining compatible with all modern web browsers:
\begin{lstlisting}
  ECDH@STRENGTH:DH@STRENGTH:HIGH:!RC4:!MD5:!DES:!aNULL:!eNULL
\end{lstlisting}

\subsubsection{Key points}
\begin{itemize}
\item Make sure server cipher setting prevails over the client's one.
\item Configure cipher suite server-side to disable insecure ciphers.
\item For the sake of interoperability, do not restrict available ciphers too much.
\item Favor ciphers that can do PFS, and cipher with longer keys. 
\end{itemize}

\subsubsection{How-to}
\paragraph{Compare two cipher lists}~

\begin{lstlisting}
  openssl ciphers ALL:-NULL |tr ':' '\n' > /tmp/1 
  openssl ciphers ALL:!aNULL:!eNULL |tr ':' '\n' > /tmp/2 
  diff /tmp/1 /tmp/2|less
\end{lstlisting}

\paragraph{Dump all ciphers supported by a service (the SSLScan tool
  is an alternative)}~
  
\begin{lstlisting}
  for c in `openssl ciphers ALL|tr ':' '\n'`; do
    echo ""|openssl s_client -cipher $c -connect www.example.net:443 2>/dev/null
  done | awk '/Cipher.*:/{print $3}'
\end{lstlisting}

\subsection{Protocols}
There are now five versions of the protocol: SSLv2, SSLv3, TLSv1,
TLSv1.1 and the latest TLSv1.2. SSLv2 is very well known 
to be insecure now and should \textit{not} be used nowadays. Of course, using 
the latest version of the protocol is desirable, but it can only be used if both
server and client support it. Most of the time it does not happen, 
which means that a server must support versions down to SSLv3 in order
to avoid blocking older clients.

Newer versions of the protocol introduce support for stronger ciphers, 
and fix various vulnerabilities that existed in SSLv3 and TLSv1.
In an ideal world, these two earlier versions of the protocol should be
phased out, but unfortunately, support for TLSv1.1 and TLSv1.2 is still
far from being universal in web browsers, therefore we need to support 
older SSLv3 and TLSv1, and work around their security flaws.

As an administrator, there are therefore very little to do with protocol
versions. Disable SSLv2, and make sure your software is recent enough to 
use TLSv1.2. If you happen to develop software that uses TLS, there are
a few things to know, though.

When using the OpenSSL library in the C language and initializing
SSL contexts, a \lstinline{SSL_METHOD} must be provided, and this 
\lstinline{SSL_METHOD}
is obtained by a set of functions that helps selecting the TLS
version protocol.  Here are the client side flavors (server side
flavors look exactly the same):
\begin{itemize}
\item \lstinline{SSLv23_client_method()}
\item \lstinline{SSLv3_client_method()}
\item \lstinline{TLSv1_client_method()}
\item \lstinline{TLSv1_1_client_method()}
\item \lstinline{TLSv1_2_client_method()}
\end{itemize}

Using \lstinline{TLSv1_2_client_method()} looks appealing, but
unfortunately this function only enables TLSv1.2, which means that
it will prevent from connecting to a server that does not support
TLSv1.2. On the other hand, \lstinline{SSLv23_client_method()}
looks undesirable because it may enable insecure SSLv2. But as the
function name does not suggest, \lstinline{SSLv23_client_method()}
 is able to negotiate the highest protocol version available, up
to latest TLSv1.2 if the server supports it. Using
\lstinline{SSLv23_client_method()} while explicitly disabling
SSLv2 (this is done using \lstinline{SSL_CTX_set_options()} with
\lstinline{SSL_OP_NO_SSLv2}) will therefore bring you the best
trade off between security and compatibility.

Another common issue is software that do not setup certificate
validation. This should be done with the following functions:
\begin{itemize}
\item \lstinline{SSL_set_verify()} or \lstinline{SSL_CTX_set_verify()} :
   these functions take a mode, which should obviously not be 
   \lstinline{SSL_VERIFY_NONE} for a client that wants to authenticate a server.
\item \lstinline{SSL_set_verify_depth()} or 
\lstinline{SSL_CTX_set_verify_depth()} :
   this sets the maximum certificate chain depth when validating the
   certification chain.
\end{itemize}

PHP programmers face the same problems. Various function accept a
\lstinline{ssl://} or \lstinline{tls://} URL as an argument. Here again,
\lstinline{tls://} looks more modern and more desirable, but in PHP
5.3.x it will translate into \lstinline{TLSv1_client_method()},
which means only TLSv1.  \lstinline{ssl://} will use
\code{SSLv23\_client\_ method()} and enable up to TLSv1.2 if
possible.  Using \lstinline{ssl://} instead of \lstinline{tls://} seems
therefore better, but unfortunately, PHP 5.3.x does not configure
OpenSSL so that \lstinline{SSLv23_client_method()} is unable to
use SSLv2, hence we have to be sure the server disabled it.

Another PHP 5.3.x issue is that by default, no certificate validation
is done. It is possible to enable it using stream context options, 
and while there, ciphers can also be specified. Here is an example,
with backward compatibility code for older PHP versions (where 
certificate validation will not be done):

\begin{lstlisting}[language=PHP]
if (function_exists('stream_socket_client') {
  $remote = sprintf("ssl://%s:%d", $host, port);
  $opts = array(
    'ssl' => array(
      'verify_peer' => TRUE,
      'verify_depth' => 5,
      'cafile' => '/path/to/ca_file',
      'ciphers' => 'ECDH@STRENGTH:DH@STRENGTH:!RC4:!MD5:!DES:!aNULL:!eNULL',
    ),
  );
  $ctx = stream_context_create($opts);
  $timeout = ini_get("default_socket_timeout");
  $stream = @stream_socket_client($remote, $errorNumber, $errorString,
                                  $timeout, STREAM_CLIENT_CONNECT, $ctx);
} else { /* Backward compatible fallback without certificate validation */
  $stream = @fsockopen('ssl://' . $host, $port, $errorNumber, $errorString);
}
\end{lstlisting}

Other languages will often face the same kind of issues. Generally 
speaking, it is a good idea to check what happens when connecting to 
a server with a self signed certificate. If it works without an error,
it means some specific code must be added to perform certificate
validation. Disabling SSLv2 and using the best protocol version is
also often a non trivial issue that should be inspected.

\section{TLS Configuration hardening in practice: Apache}

We will now look at how to harden TLS configuration for 
a few services: Apache, Sendmail, Dovecot, OpenVPN... Since we talk about
examples, we will of course not cover every software available, but
once you know what to look at, it is easy to do the same for another
program. Our first target will be Apache.

First let us look at what we already covered:
\begin{itemize}
\item Make sure your Apache is recent enough to support TLSv1.2 and ECDH
  Latest Apache 2.2.x or Apache 2.4.x will do it.
\item Generate a RSA private key at least 2048 bits long
\item Make sure your certificate is signed by a CA known by browsers
\item Use the following configuration for Apache
\begin{lstlisting}[language={}]
 SSLProtocol all -SSLv2
 SSLHonorCipherOrder On
 SSLCipherSuite ECDH@STRENGTH:DH@STRENGTH:HIGH:!RC4:!MD5:!DES:!aNULL:!eNULL
\end{lstlisting}
\end{itemize}

Removing SSLv3 (using \lstinline{SSLProtocol all -SSLv2 -SSLv3}) 
and triple DES (using \lstinline{!3DES} in \code{SSL\-CipherSuite})
would be desirable, but it may lock out some 
older clients. As older devices will get replaced, this will become 
possible without any drawback. Logging protocols and ciphers selected
by your clients is the only way to know if time has come to phase out 
SSLv3 and triple DES.

Additionally, if you do not serve clear text content, let the client know
it by using the Strict Transport Security HTTP header. Once it has been
received by a browser, it will force HTTP over TLS even if the user 
follows a \lstinline{http://} URL setup on a malicious site. 
This thwarts possible MITM attacks. Here is how to set it up on Apache:
\begin{lstlisting}[language={}]
 Header set Strict-Transport-Security "max-age=15768000"
\end{lstlisting}

There is also an older mechanism specific to cookie protection.
Cookies are used to maintain sessions on web applications, and they
are therefore authentication credentials that should not leak. When
setting a cookie, one can set the \lstinline{secure} flag, which
means the cookie must be sent only over TLS protected connexions.
Although redundant for cookie protection, it does not hurt to setup
both \lstinline{secure} cookies and \lstinline{Strict-Transport-Security}.
Cookies can also have the \lstinline{httpOnly} flag, which means
the JavaScript environment cannot access them. It is a good idea
to use it, as it reduces the attack surface on session cookies.
Most of the time this requires modifying application code, but
there are situations where it can be achieved by the administrator.
For instance, applications using PHP sessions can get both
\lstinline{secure} and \lstinline{httpOnly} cookies by setting
\lstinline{session.cookie_secure} and \lstinline{session.cookie_httponly}
to \lstinline{On} in \lstinline{php.ini}.

Next, test your setup. There is a very useful SSL server test \cite{10} on the
web for that, thanks to Qualys SSL Labs. Run the test and it will point most
mistakes in your TLS configuration. If you used the configuration suggested
in this paper, you will probably have a bad mark. Why? There can be two
kind of issues.

First possible problem, you read this paper a long time after its
publication, and the sample configuration is now insecure. You will 
have to adjust it with the help of Qualys SSL server test. If your
certificate key length or hashing algorithm is now weak, improve it
with better settings. If the tests says a TLS version is now insecure, 
adjust \lstinline{SSLProtocol} to remove it. If you have problem with ciphers, 
look at the test results for browser handshake simulation. Check what 
cipher is picked by each browser, and adjust 
\lstinline{SSLCipherSuite}.

Once you fixed certificate, protocol and cipher, you may still get a bad
mark, despite the TLS configuration being pretty good. This is because 
the web is a quite complex environment, which makes some web-specific 
attacks against TLS possible, and hence we need to workaround them.

\subsection{Mitigating the CRIME attack}

The CRIME \cite{11} attack belongs to the chosen plain text class of attacks. It
allows the attacker to extract authentication cookies from the TLS-protected
data stream, which allows session hijacking. The CRIME attack can happen if: 
\begin{enumerate}
\item the browser is tricked into sending data controlled by the attacker
\item the attacker can observe the server response
\item TLS compression is enabled
\end{enumerate}

The first two conditions can happen if the attacker controls content 
from the served pages, which can happen when the server has Cross
Site Scripting (XSS) vulnerabilities. The other situation where it
may happen are cross-site requests, where a malicious web site 
tricks the browser into sending data to the attacked web site.

The only CRIME workaround is to work on the third condition, 
and to disable TLS compression. this is the default in recent Apache 
releases, as in recent browsers. On earlier Apache releases, the 
following configuration directive lets you disable it:
\begin{lstlisting}[language={}]
  SSLCompression  off
\end{lstlisting}

Disabling TLS compression does not affect performance a lot, since
on a well configured server, HTTP compression is enabled. But as we 
will see in the next section, it also brings problems.

\subsection{Mitigating the BREACH attack}

The BREACH \cite{12} attack is similar to CRIME. Here the offending feature is
not TLS compression but HTTP compression. A quick fix is to disable 
HTTP compression, but that could have a bad impact on performances. 

A nice solution is to disable HTTP compression for requests coming from
another web site. That will not help if the attacker uses a XSS 
vulnerability, but there are many other ways to attack your site in 
that scenario, hence we will assume you already fixed it.

On the other hand, that prevents an attacker from running a BREACH
attack from another web site. This workaround can be implemented this
way in Apache, assuming your server name is 
\lstinline{www.example.net}:

\begin{lstlisting}[language={}]
        # BREACH mitigation
        SetEnvIfNoCase Referer .* self_referer=no
        SetEnvIfNoCase Referer ^https://www\.example\.net/ self_referer=yes
        SetEnvIf self_referer ^no$ no-gzip
\end{lstlisting}

\subsection{Mitigating the BEAST attack}

The BEAST attack works on SSLv3 and TLSv1, against a class of cipher
known as block ciphers (most of the time they have CBC, like Cipher
Block Chaining, in their names). The obvious fix is to disable block 
ciphers,  but that leaves us with the following ciphers available:
\begin{description}
\item[GCM] (Galois/Counter Mode) ciphers
\item[RC4] Rivest Cipher v4
\end{description}

GCM ciphers are only available with TLSv1.2, which means it does 
not help since we look for a mitigation against an attack on
SSLv3 and TLSv1.

RC4 could help, and at some time it was advised to craft a 
\lstinline{SSLCipherSuite}
setting that favored GCM ciphers, then RC4, so that newer browsers could
pick GCM and older would use RC4. Here is an example:
\begin{lstlisting}[language={}]
ECDH@STRENGTH:DH@STRENGTH:-SHA:ECDHE-RSA-RC4-SHA:HIGH:!MD5:!DES:!aNULL:!eNULL
\end{lstlisting}

Unfortunately, RC4 itself was discovered to be weaker than previously 
thought \cite{13}, leaving system administrators with the choice between two 
evils: be vulnerable to a RC4 attack that requires hours of exchanges, or 
be vulnerable to BEAST attacks.

The former alternative may look more desirable, but web
browsers made progress at defending against BEAST, either by supporting
TLSv1.1 or TLSv1.2, or by implementing a hack called the 1/n-1 split,
that makes TLSv1 CBC ciphers not vulnerable to BEAST. On the other hand, 
there is no way to defend against attacks on RC4, and they are likely to 
get more efficient at times goes.

Favoring RC4 hence improves the security for browsers still vulnerable
to BEAST, but it decreases it for browser that implemented BEAST
workarounds and would be pushed to pick a weak RC4 cipher instead of 
for instance AES256.

This is the reason why today it makes sense to disable RC4 and favor
BEAST-vulnerable, PFS-capable block ciphers. This is what we did
in the suggested SSLCipherSuite in this paper, but there is no broad
consensus on this: Some web site still prefer to favor RC4. This
is the case for Google as of may of 2014. It makes some sense because
today, an attack on RC4 still requires a lot of exchanges.

Despite BEAST mitigation being widely available, you may have clients
that have not been updated and are still vulnerable. If you chose
a server configuration that promotes strong ciphers but leaves older
clients vulnerable to BEAST (i.e.: you did not favor RC4), you
may want to detect vulnerable clients, and report the problem to
the user in order to push for a browser update. The author of this
paper wrote an Apache module for that \cite{14}. Qualys SSL Labs also has
a more feature-rich browser fingerprinting Apache module that can
detect a BEAST vulnerable browser \cite{15}.

\subsection{Session renegotiation}

In 2009, a vulnerability was discovered in a feature called session 
renegotiation \cite{16}. As the name suggests, it allows the change
of TLS session parameters. For instance, you could have a part
of a web site that wants the client to authenticate using an 
X.509 certificate, while the remaining of the site could be 
accessed without a certificate.

When the client connects to the web server and negotiate the TLS 
session, the server does not know the requested URL yet. In order
to enforce a per-URL policy for requiring client certificate, the
server needs to renegotiate the TLS session after the URL has been
sent. This is that exact feature that has been under attack. The
TLS renegotiation procedure has been updated in the TLS protocol 
to work around the problem.

The only reasonable fix here is to upgrade the server so that
it support secure TLS session renegotiation. This will not be
backward compatible with older browsers that have not been 
updated. If you need to support them, you will have to give 
up on per-URL policy, and use multiple virtual servers using
different IP addresses. That way Apache knows what to do before
the TLS handshake is done, and TLS renegotiation is not needed 
anymore.

\subsection{Fixing Heartbleed}

Heartbleed is the TLS attack that got the most press coverage, thanks to 
the communication efforts of the researchers that discovered it: a good
name, a dedicated web site, and perhaps a first time for a computer
vulnerability, a dedicated logo. And unfortunately, the vulnerability 
itself is extremely severe, making legitimate the efforts to get it
picked up by mainstream press.

As opposed to protocol vulnerabilities like CRIME, BEAST and BREACH,
Heartbleed is an implementation vulnerability : just a programming
error in OpenSSL that introduces a vulnerability in a TLS extension 
called TLS Heartbeat. Other TLS implementations such as GnuTLS or 
Mozilla NSS are not vulnerable, neither are OpenSSL 0.9.x releases
and earlier, which did not have the TLS Heartbeat functionality.

The bug is a buffer overflow that allows an attacker to read 64 kB 
of memory from the server, which can include various data recently 
used, like session cookies, passwords, database access credentials, 
and in the worst case, the server private key (the only case where
the private key remains safe while the server is vulnerable is when 
it is stored in a hardware security module, because it never appears
in memory).

Buffer overflow mitigation techniques such as Address Space Layout
Randomization (ASLR) are not effective against Heartbleed because
OpenSSL implemented its own memory management on top of libc's one,
for performance's sake. That situation is controversial since it is
not granted that OpenSSL's memory management layer is faster than 
libc's one on any system. Even on systems where it is, this is a
problem that should be fixed in libc itself, and not worked around
in OpenSSL.

Even more unfortunate is the fact that TLS Heartbeat is an optional
extension that is almost useless. It is used to keep connexions
alive in Datagram TLS (DTLS), which is TLS over UDP. When TLS runs
on top of TCP, which is the case for HTTP over TLS, the problem
solved by TLS Heartbeat does not exist.

Heartbleed is only an implementation vulnerability, 
which means it will be fixed by just
upgrading OpenSSL to at least version 1.0.1g, or by rebuilding it without
TLS Heartbeat enabled (that is, using 
\code{-DOPENSSL\_NO\_ HEARTBEATS} option
to the C compiler). Since the vulnerability may have allowed private
keys to be extracted, they must be replaced, TLS certificates must
be renewed and the older certificate should be revoked. Then all user 
passwords should be changed. All that work is painful, but probably 
necessary.

There have been some controversy on how easily the server private
key could be extracted. The vulnerability only allows reading 64
kB of data, and most of the time the private key is out of reach,
but it has been shown that it can be reliably obtained just after
a server restart \cite{17}. Reliable or not, a private key-leak
vulnerability is still a major concern, as one leak is enough
to compromise the private key forever. And moreover, the vulnerability
has existed for two years before being disclosed and fixed, hence
it cannot be excluded that someone knew it and used it during that
time frame.

\subsection{Ineffective certificate revocation vulnerabilities}

There is a dark corner of Heartbleed recovery that has not been 
covered a lot: many web browsers never check for certificate
revocation, and even if they do, there are situations where they
fail to do it and still allow the connexion to proceed (WiFi hotspot
is a situation where it happens). As a result, if an attacker 
stole a private key, he is still able to use it with the revoked 
certificate to perform Man in the Middle attacks. This means an effort
will have to be done to improve certificate revocation enforcement.

There are two widely supported methods to check for certificate
revocation. The first one is Certificate Revocation List (CRL), a
CA-signed list of revoked certificate serial numbers, which is
published at a well known URL carved into the CA certificate. This
methods obviously does not scale, as the file length only increases
as certificates are revoked. And it also requires a network connexion.

The other major approach is Online Certificate Status Protocol (OCSP),
a protocol that lets the browser ask for the revocation status of a
particular certificate. It scales much better than CRL, but it still
requires a network connexion to be used.

Google introduced an offline mechanism in Chrome, which is called 
CRL sets. The idea is that Chrome gets CRL with its automatic updates,
removing the need for a network connexion when a certificate revocation
status is checked. Unfortunately this mechanism suffers a scaling problem
even worse than CRL, as CRL sets should list revoked certificate serial 
numbers for all the CA known to the browser. This would be huge, and 
as a result, Google only provide CRL sets for a subset of the installed CA
\cite{18}, leaving the user vulnerable without any clue about which web site
is protected by CRL sets and which one is not.

The problem could be addressed correctly by OCSP stapling, which is
a TLS extension that enable the TLS server to acquire a certificate 
validity proof from the CA so that it can be presented to the client 
during the TLS handshake. In other words, the TLS server acts as an 
OCSP proxy for the client. This both solves scalability problem (except
perhaps for the network bandwidth consumed by OCSP services) and the
network access requirement: a not-yet-authenticated host on a WiFi hotspot
is able to assess the authentication server's certificate revocation
status. But unfortunately, OCSP stapling requires client and server
support, and it is not yet widely supported.

\subsection{Other attacks}

There are other attacks against TLS for which Qualys SSL server
test will not warn you about. The padding oracle attack was discovered
in 2002, and it was worked around for a long time by all implementations,
but a new usable variant called Lucky 13 was published in 2013.
Like BEAST, Lucky 13 targets CBC ciphers, which means it can be
mitigated by using RC4 (but this trades a vulnerability for another one), 
or by using TLSv1.2, which brings GCM ciphers. There is also a Lucky 13 
implementation workaround for CBC ciphers, obtained by introducing time 
variations in the algorithms. But both GCM and CBC with time variations
are not implemented by all clients, and there is no way for the server
to know if Lucky 13 mitigation for CBC is done on the client or not.

Fortunately, Lucky 13 needs the attacker to temper with TLS traffic,
which means it must be in a position to act as a Man in the Middle
(MITM). And since it is a timing dependent attack, it requires a
network connexion with low jitter (jitter is the variation of
latency), and a lot of requests to succeed and recover a session
cookie. The risk is hence low for now. It can be lowered by limiting
session cookie lifetime, or even better, by binding it to the number
of requests done.

\subsection{Client vulnerabilities}

As noted earlier, some TLS vulnerabilities require client fixes.
This is the case for TLS session renegotiation, and BEAST, thought
the 1/n-1 split workaround in CBC ciphers. Lucky 13 can also be spared
by implementing timing variation for CBC ciphers. If we assume the
client has not been updated, we can have server-side mitigation,
by avoiding TLS session renegotiation, or by using RC4 (but as already
said it is a controversial choice, since RC4 is vulnerable to other attacks), 
or GCM ciphers available in TLS 1.2 (which is still not supported by
many clients). However there are other threats against clients, for 
which no server-side mitigation is available.

The Apple TLS stack suffered a security failure because of a duplicate
"goto fail" statement in the source code. It allowed easy hijacking
of TLS session for TLS up to TLSv1.1 and ciphers using ECDHE or DHE
(these are the ciphers required to support Perfect Forward Secrecy).
This caused a widespread vulnerability on iOS, and vulnerabilities in
Safari and Apple Mail on MacOS X. The only decent workaround is to 
upgrade the OS. On MacOS X, software that bring their own TLS 
implementation, such as Firefox and Thunderbird, remained safe.

As for the BEAST vulnerability, identifying the browsers vulnerable 
to the "Apple goto fail" vulnerability is a good idea, since it
allows users to be notified. This can be done by running a dedicated
Apache linked with a patched OpenSSL, as described by Qualys SSL labs'
Ivan Ristic \cite{19}. 

If you have a corporate intranet start page, it is a good place to run
Multiple browsers vulnerabilities tests. This can easily be achieved
by including \lstinline{<img>} tags for 1x1 images on a test server. If a 
vulnerability is identified, a site-wide cookie can be be set, and 
the corporate intranet start page can use it to warn the user.

While there, you can also test if the browser accepts self-signed
certificates, which would allow anyone to perform easy Man in the
Middle attacks. Some mobile browser fail to perform this basic
check, which means their TLS implementation does not bring much
more security than clear text communications.

Key Points
\begin{itemize}
\item Use up to date server software
\item Test your TLS-enabled web server with Qualys SSL server test
\item Make sure TLS compression is disabled
\item Disable HTTP compression for requests incoming from other sites
\item Until the day TLSv1.2 support becomes widespread, choose between
  dangers of RC4 and CBC ciphers.
\item Assess browser vulnerabilities and notify users so that they know they
  need an upgrade
\item Protect session cookies with the 
\lstinline{secure} and \lstinline{httpOnly} flags
\end{itemize}

\section{Hardening other protocols}

Once your web servers are secure, it is time to look at other protocols:
here are a few obvious examples: POP, IMAP, SMTP, LDAP, OpenVPN.

The good news is that many attacks possible on HTTP over TLS cannot be
transposed to other protocols. This is because there is no way to trick the
client into sending chosen data. POP requests cannot bounce from an
external server to yours, and no JavaScript can inject data into a
SMTP session.

This is true for all above mentioned protocols. It means attacks on 
CBC ciphers such as BREACH, CRIME or BEAST cannot happen here. The only 
thing we will have to worry beyond certificates, protocols and ciphers 
are Lucky 13 and RC4 vulnerabilities.

Dealing with RC4 vulnerability is simple, we can just disable it. This is an 
easy choice here since we do not choose between RC4 attacks and BEAST. 
On the other hand, protecting against Lucky 13 is not easy. We cannot
enforce the use of TLS 1.2 GCM ciphers, as many clients do not support
them, and we do not want to use RC4. We may assume timing variant workarounds
to be implemented in the client, we we have no way to assess it is the case.
Unfortunately, this client workaround, and the difficulty to run a 
Lucky 13 attack, are our only solutions against Lucky 13 until the 
day TLS 1.2 is universally supported.

\subsection{Dovecot}

Here is a sample dovecot TLS settings:

\begin{lstlisting}[language={}]
  ssl = yes
  ssl_ca = </etc/openssl/certs/ca-chain.crt
  ssl_key = </etc/openssl/private/server.key
  # Server certificate with CA chain included
  ssl_cert = </etc/openssl/certs/server-bundle.crt
  # No need to disable SSLv2, it is done by default
  ssl_prefer_server_ciphers = yes
  ssl_cipher_list = ECDH@STRENGTH:DH@STRENGTH:HIGH:!RC4:!MD5:!DES:!aNULL:!eNULL
  disable_plaintext_auth = yes
\end{lstlisting}

Note that ECDHE is available starting with Dovecot 2.2.6. As for web server,
supporting it is important because it will increase the amount of
Perfect Forward Secrecy capable clients.

SSLv2 is supposed to be disabled by default, but it does not hurt to 
check that:
\begin{lstlisting}
  openssl s_client -ssl2 -connect server:993
\end{lstlisting}

\subsection{Sendmail}

Many TLS features in Sendmail need to be enabled at compile time. This 
is achieved by filling the \lstinline{site.config.m4} before the build:

\begin{lstlisting}[language={}]
 # enable STARTTLS
 APPENDDEF(`conf_sendmail_ENVDEF', `-DSTARTTLS')
 APPENDDEF(`conf_sendmail_LIBS', `-lssl -lcrypto')
 # enable _FFR_TLS_1, for CipherList directive
 APPENDDEF(`conf_sendmail_ENVDEF', `-D_FFR_TLS_1')
 # enable _FFR_TLS_EC, for ECDH support, requires sendmail 8.14.8 or higher
 APPENDDEF(`conf_sendmail_ENVDEF', `-D_FFR_TLS_EC')
\end{lstlisting}

If your Sendmail was obtained as a binary package, the following command
let you know what is built inside (look for \lstinline{STARTTLS}, 
\lstinline{_FFR_TLS_1} and \lstinline{_FFR_TLS_EC}):
\begin{lstlisting}
 sendmail -d0.13 < /dev/null 
\end{lstlisting}

\lstinline{_FFR_TLS_EC} was contributed by the author of this paper. It is really new 
and is not likely to be built in a binary package in the months following
the publication of this article. Also note that FFR stands for "For Future 
Release", which suggests that the feature may become default in future. 

Once you have the correct options, either because you already had them, or
because you made a custom Sendmail build, you can use the following options 
in \lstinline{sendmail.cf}:

\begin{lstlisting}[language={}]
 O CACertPath=/etc/openssl/certs/
 O CACertFile=/etc/openssl/certs/ca-chain.crt
 O ServerCertFile=/etc/openssl/certs/server.crt
 O ServerKeyFile=/etc/openssl/private/server.key
 O DHParameters=/etc/openssl/certs/dh.pem
 O CipherList=ECDH@STRENGTH:DH@STRENGTH:HIGH:!RC4:!MD5:!DES:!aNULL:!eNULL
 O ServerSSLOptions=+SSL_OP_NO_SSLv2 +SSL_OP_CIPHER_SERVER_PREFERENCE
 O ClientSSLOptions=+SSL_OP_NO_SSLv2
\end{lstlisting}

Recent Sendmail will not need the 
\lstinline{DHParameters} option, but it is worth
a few words. ECDHE and DHE ciphers need DH parameters to operate. Some
softwares are able to auto-generate them internally, while others will
require the administrator to generate a DH parameter file and specify 
it in the configuration. The file can be obtained by running the following
command:
\begin{lstlisting}
 openssl dhparam 2048 > /etc/openssl/certs/dh.pem
\end{lstlisting}

\subsection{OpenLDAP}

TLS support in OpenLDAP has been good for a long time, hence you are
not likely to have missing features. 

There are two cases here: the TLS service available to clients, and the
TLS connexions between master server and replicas. The former needs to
be permissive enough so that any client can connect, but the later
is a good candidate for hardening since it only needs to support
connexions from replicas using recent OpenLDAP releases.

Here is a hardened configuration for an OpenLDAP master: 

\begin{lstlisting}[language={}]
  TLSCertificateFile      /etc/openssl/certs/server.crt
  TLSCertificateKeyFile   /etc/openssl/private/server.key
  TLSCACertificateFile    /etc/openssl/certs/ca-chain.crt
  TLSDHParamFile          /etc/openssl/certs/dh2048.pem
  TLSCipherSuite          ECDH:DH:!RC4:!SHA:!MD5:!DES:!aNULL:!eNULL
  TLSVerifyClient         allow
  TLSCACertificatePath    /etc/openssl/certs
  TLSCRLCheck             all
\end{lstlisting}

\lstinline{TLSCACertificateFile}, \lstinline{TLSVerifyClient}, 
\lstinline{TLSCACertificatePath} and \lstinline{TLSCRLCheck}
are useful if you allow clients to authenticate using certificates, which 
is a good idea for LDAP replicas connecting to a LDAP master (more on 
that later).

Note that \lstinline{!SHA:!MD5} in the CipherSuite will disable all ciphers 
from TLSv1.1 and earlier : \lstinline{SHA} in a cipher specification stands for
SHA1, and only TLSv1.2 brings ciphers using hash algorithm 
using SHA2. 

The specification may be odd, but it is just because it is an evolution
from the previous cipher suite we used before. The same result could be
achieved with something perhaps more easy to understand such as:
\begin{lstlisting}[language={}]
   ECDH:DH:!TLSv1:!SSLv3:!aNULL:!eNULL
\end{lstlisting}

On the replica side, the cipher suite needs to allow more clients to
connect. You can reuse the cipher suite we used for the web, or 
if your LDAP clients are good enough, you can force PFS usage:

\begin{lstlisting}[language={}]
  TLSCertificateFile      /etc/openssl/certs/server.crt
  TLSCertificateKeyFile   /etc/openssl/private/server.key
  TLSCACertificateFile    /etc/openssl/certs/ca-chain.crt
  TLSDHParamFile          /etc/openssl/certs/dh2048.pem
  TLSCipherSuite          ECDH@STRENGTH:DH@STRENGTH:!RC4:!MD5:!DES:!aNULL:!eNULL
  TLSVerifyClient         allow
  TLSCACertificatePath    /etc/openssl/certs
  TLSCRLCheck             all
\end{lstlisting}

OpenLDAP replicas can also have a TLS setup for their client-side, 
in chain overlay and syncrepl configuration. Here is a configuration
sample:

\begin{lstlisting}[language={}]
syncrepl rid=17
    provider=ldap://ldap-master.example.net
    type=refreshAndPersist
    searchbase="dc=example,dc=net"
    starttls=critical
    schemachecking=off
    sizelimit=unlimited
    retry="3 1 10 2 60 +"
    bindmethod=sasl
    saslmech=EXTERNAL
    tls_cert=/etc/openssl/certs/server.crt
    tls_key=/etc/openssl/private/server.key
    tls_cacert=/etc/openssl/certs/ca-chain.crt
    tls_reqcert=demand
    tls_cipher_suite=ECDH:DH:!RC4:!SHA:!MD5:!DES:!aNULL:!eNULL
    tls_cacertdir=/etc/openssl/certs
    tls_crlcheck=all
\end{lstlisting}

Both the master and replica configuration here include directives to enforce
CRL checking. OpenSSL is a bit odd here and deserves some explanations. If
CRL checking is enabled using 
\lstinline{TLSCRLCheck} (\lstinline{tls_crlcheck} in syncrepl 
configuration), then OpenSSL will look for a CRL file in the directory 
specified by \lstinline{TLSCACertificatePath} (\lstinline{tls_cacertdir}
 in syncrepl configuration).
The name of the CRL file is of the form 
\lstinline{$hash.r0}, where \lstinline{$hash} is
obtained by the \lstinline{openssl crl} command:
\begin{lstlisting}
    openssl crl -hash -noout -in ca.crl
\end{lstlisting}

This means that once you obtained the \lstinline{ca.crl} file, you must
install it in the directory specified by 
\lstinline{TLSCACertificatePath} (\lstinline{tls_cacertdir}), and
run the following command:
\begin{lstlisting}
    ln -s ca.crl `openssl crl -hash -noout -in ca.crl`.r0
\end{lstlisting}
   
If you manage your own internal CA, which is a good idea for LDAP
certificates, the CRL file can be obtained from the CA key and
certificate by using this command:
\begin{lstlisting}
    openssl ca -keyfile ca.key -cert ca.crt -gencrl -out ca.crl
\end{lstlisting}

\subsection{Webmail software}

We dealt with web and mail servers, but not in the middleware that can
sit between them, which is known as webmail. Many software are available, 
but a close inspection of a few of them (SquirrelMail, RoundCube, ImapProxy)
has shown a few deficiencies: no certificate validation was done, no effort
was done to push TLSv1.2 usage if available, and cipher suites were not
configurable. 

While the two latter issues can be acceptable, the lack of certificate 
validation can be a real problem, as it allows easy Man in the Middle 
attacks. Often the risk is neglected because mail and web servers sit
close to each other in the same data center, but that may not always be
the case.

Patches have been submitted and accepted in the three above mentioned
software to improve the situation but as of today formal releases 
including the patches have not been done. Here are the minimum
version requirement  and configuration snippet to get support up to 
TLSv1.2, ECDHE ciphers, CA validation, and configurable cipher suites:

\paragraph{ImapProxy > 1.2.7} (Use SVN version)
\begin{lstlisting}[language={}]
	tls_ca_file  /etc/openssl/certs/tcs-chain.crt
	tls_ciphers ECDH:DH:!RC4:!SHA:!MD5:!DES:!aNULL:!eNULL
	tls_verify_server true
	force_tls true
\end{lstlisting}

\paragraph{Squirrelmail > 1.4.22} (Use 1.4.23-svn or 1.5.2-svn versions)
\begin{lstlisting}[language=PHP]
	$use_smtp_tls = true;
        $smtpOptions['ssl']['verify_peer'] = true;
        $smtpOptions['ssl']['verify_depth'] = 3;
        $smtpOptions['ssl']['cafile'] = '/etc/openssl/certs/ca-chain.crt';
        $smtpOptions['ssl']['ciphers'] = 
	    'ECDH:DH:!RC4:!SHA:!MD5:!DES:!aNULL:!eNULL';
        $smtp_stream_options = $smtpOptions;
\end{lstlisting}

\paragraph{Roundcube $\geq$ 1.0rc}~
\begin{lstlisting}[language=PHP]
	$rcmail_config['smtp_conn_options'] = array(
	  'ssl' => array(
	    'verify_peer' => TRUE,
	    'verify_depth' => 3,
	    'cafile' => '/etc/openssl/certs/ca-chain.crt',
	    'ciphers' => 'ECDH:DH:!RC4:!SHA:!MD5:!DES:!aNULL:!eNULL';
	  ),        
	);
\end{lstlisting}
 
There are a few points to note:
\begin{itemize}
\item SquirrelMail and RoundCube certificate validation and cipher
  specification use PHP context options, which are only available in PHP 5.
\item We assume ImapProxy usage here: IMAP configuration for SquirrelMail
  and Roundcube is therefore not using TLS. 
\item We assume the mail server TLS setup is under our control and that we
  can enforce a restrictive cipher suite, with only PFS-enabled TLSv1.2.
\end{itemize}

\subsection{OpenVPN}
The case of OpenVPN could be straightforward: it enjoyed very good TLS 
support for a long time, and we could assume clients are up to date. 
This would let us enforce the use of TLSv1.2 GCM ciphers. Here are the 
TLS related options in server configuration:
\begin{lstlisting}[language={}]
  ca /etc/openssl/certs/ca-chain.crt
  cert /etc/openssl/certs/server.crt
  key /etc/openssl/private/server.key
  dh /etc/openssl/certs/dh2048.pem
  tls-cipher ECDH:DH:!RC4:!SHA:!MD5:!DES:!aNULL:!eNULL
  client-cert-not-required
\end{lstlisting}

But unfortunately there are many outdated clients that will not support
such a setup. TLSv1.2 is not available in OpenVPN up to 2.3.2, and
MacOS X's package GUI for OpenVPN, which is known as Tunnelblick, is
bundled with various versions of OpenVPN (as of may 2014, latest Tunnelblick
ships with OpenVPN version 2.2.1, 2.3.2, and 2.3.4). The preference panel 
lets the user choose the OpenVPN version, and if it is not set to 2.3.4, or if
Tunnelblick is too old to include 2.3.4, the client will not work with a
TLSv1.2 only configuration. The setup below is more compatible by allowing
CBC ciphers from protocols versions prior to TLSv1.2:
\begin{lstlisting}[language={}]
  ca /etc/openssl/certs/ca-chain.crt
  cert /etc/openssl/certs/server.crt
  key /etc/openssl/private/server.key
  dh /etc/openssl/certs/dh2048.pem
  tls-cipher ECDH:DH:!RC4:!MD5:!DES:!aNULL:!eNULL
  client-cert-not-required
\end{lstlisting}

\subsection{More protocols}

There may be other protocols to look at. To name a few, RADIUS configuration
for EAP-TTLS in 802.1x authentications, SIP, or any TCP service wrapped 
into Stunnel. 

This latter utility is worth mentioning, as it allows any TCP stream to
be protected by TLS. If you encounter a software with broken TLS that
cannot be fixed, disabling its TLS support and replacing it with Stunnel
is a good idea: the broken software listens for plain text traffic on 
localhost, and Stunnel offers the TLS-protected service over the network, 
and relays it to the broken software. The Stunnel configuration will looks
like this:
\begin{lstlisting}[language={}]
  chroot = /var/chroot/stunnel/
  setuid = stunnel
  setgid = stunnel
  pid = /stunnel.pid
  debug = 0
  CAfile = /etc/openssl/certs/ca-chain.crt
  sslVersion = TLSv1
  ciphers = ECDH@STRENGTH:DH@STRENGTH:HIGH:!RC4:!MD5:!DES:!aNULL:!eNULL
  
  [https]
  # contains CA chain and certificate
  cert = /etc/openssl/private/cert-bundle.crt
  key =  /etc/openssl/private/server.key
  accept = 443
  # protected service is told to listen on port 127.0.0.1:8080
  connect = localhost:8080
\end{lstlisting}

Note that such a setup is also useful for debugging TLS-protected protocols. 
Network sniffers such as tcpdump are of little interest when facing TLS
protected streams. But if you install a Stunnel relay, it is possible
to capture the clear text protocol by listening on the local interface:
\begin{lstlisting}
  tcpdump -ni lo0 -s0 -X 'port 8080'
\end{lstlisting}

\section*{Conclusions}
We had a look at TLS and how to harden it against the usual attacks
against it for various open source software. The method is always the
same: make sure software is up to date, disable SSLv2, choose a strong
cipher suite that still allow older clients to connect, and test the 
result. 

Server testing should make sure that insecure protocols versions and
ciphers are disabled. Proper certificate validation should also be 
checked, as this step is sometimes missing in some software. It should
be done for clients, and for servers that perform client authentication
using X.509 certificates. Tests may spot broken software. If it cannot 
be fixed, using an up-to-date Stunnel as a relay is an option to obtain 
a properly hardened TLS service.

The final words could be an emphasis over the TLS flaws that cannot
be worked around. Only TLSv1.2 with recent ciphers is safe from any
known attack, and since some clients do not support it, the
administrator often needs to choose between two kind of troubles:
on one hand, using CBC ciphers that are subjects to various attacks
that may be fixed in the client (BEAST, Lucky 13), and/or difficult to
perform (Lucky 13). On the other hand, adopt the weak RC4, which is
vulnerable to attacks difficult to perform, but with no fix available,
whether on client or server. No consensus exists here, and as various actors
made different choices, the author of this paper chose CBC.

Unfixed TLS flaws also exist in older clients and servers, and will 
certainly happen again in the future. Even TLS-unrelated flaws can 
have an impact on TLS, for instance when a server private key is
compromised because a cracker got root access on a server, or managed
to derive it from the public key. In such a scenario, all TLS traffic
that has been captured and stored in the past can be deciphered, except 
if a PFS cipher was used.

\section*{Acknowledgments}
Thanks to Jean-Yves Migeon, Marc Dreyfus and Fredrik Pettai for reviewing 
this paper, and to Florence Henry for her help with \LaTeX.


\end{document}